\documentclass[prb,aps,superscriptaddress,endfloats, showpacs]{revtex4}

\begin {document}

\title {Confined gravitational waves for chiral matter with heat}

\author {I. Bulyzhenkov-Widicker} 
 \affiliation 
 {the Institute of Spectroscopy RAS, Troitsk, Moscow reg. 142092, Russia}

\affiliation{Department of Physics, University of Ottawa,  \\ 150,  Louis-Pasteur, Ottawa, Ontario K1N 6N5, Canada}

\bigskip
\begin {abstract}
{The GR wave self-heating of geodesic massive bodies with constant thermo-gravimechanical energies increases the brightness-to-charge ratio along spiral radial transitions in the energy-to-energy gravitation. Paired confined gravitons locally warm accelerated matter that suggests the thermodynamical origin of electromagnetic outbursts with oscillating Wien's displacements.  Damping of orbital periods by chiral GR waves is more efficient for neutron stars around giant companions than for  binary pulsars. }
\end {abstract}
 \bigskip

\pacs{04.20.Cv}
\maketitle

\bigskip

Metric formalism for the geodesic motion, initiated by Einstein and Grossmann \cite{Ein}, is not presently unified with irreversible heat generation and black-body gamma radiation.  Electromagnetic (EM) wave losses change gravitational charges in Mach-Einstein's energy-to-energy gravitation  \cite {Bul}. This reading of  General Relativity (GR) predicted the  one meter per century increase of the Earth-Sun distance.  Energy-driven geometry  also suggested to count internal (thermodynamical, non-metric) energy N  next to mechanical, $K = mc^2/{\sqrt {1-v^2c^{-2}}}$, and gravitational, $U_o$, parts of the passive GR charge, $P_o = K {\sqrt {g_{oo}}}$, in E${\ddot {\rm o}}$tv${\ddot {\rm o}}$s-type laboratory tests of non-empty energy space for flatspace gravitation in question.

 By neglecting EM radiation, one may request the strict energy conservation,
 \begin {equation}
  {\cal E} \equiv  N + K {\sqrt {g_{oo}}} =  N + K + U_o = const, 
  \end {equation}
   of probe termodynamical bodies under the free motion in static gravitational fields ($\partial_t g_{oo} = 0$).   Then a possible heat production, $d<N>/dt \geq 0$, in free bodies with internal frictions or self radiation-absorption processes can remove them from GR's rosette cycles for idealized point bodies  without internal degrees of freedom.   
      The gravito-thermodynamical balance (1) for the passive gravitational charge ${\cal E}$ 
      and the virial theorem for a finite non-relativistic motion, $d<K+U_o>\ \approx - d<K>$,  
      can relate the orbital heat release to averaged changes of potential and kinetic energies,
      $d<N>  = -\ d<K+U_o>\ \approx +\ d <K>$.
Body's kinetic (translation) and internal energies share the equipartition split of the potential energy gain, $d<U_o> / dt  < 0$ in weak central fields, for example, where  $ U_o = {\cal E}r_o/r  << 1 $ and $r_o = G {\cal E}_{_M}/c^4 << r$. 
A Keplerian cyclic motion in such fields corresponds to negligible generation of heat by slowly rotating bodies. However, the nonlinear wave intensity of accelerated gravitational charges is to be proportional to $m^2$. Therefore, the spiral motion of massive  stars can be accompanied by very intensive geodesic self-heating (up to nuclear fusion temperatures because surface cooling through electromagnetic waves is not very efficient for massive radial bodies). Without (chiral) transformations of GR gravimechanical energies into heat and further irreversible EM losses, ideal GR charges (without self-heating mechanisms) would stay on initial cyclic trajectories without observed evolution options for real thermodynamical matter.

The strict constancy of geodesic energy-charges may be required only in an energy-to-energy gravitational theory without the EM decay/pumping of the probe GR charge. Surface temperature of small astronomical bodies is mainly balanced by inward energy fluxes, reflection, emissivity, and the Stefan-Boltzmann constant for outward gamma radiation. The bulk gravitational self-heating of a massive geodesic body can contribute to its surface  temperature. Quasi-relativistic neutron stars on prolonged elliptic orbits around giant massive centers should exhibit explosive surface temperature jumps and, therefore, periodic EM outbursts. One could relate the known gamma outbursts, for example \cite {Val},  to peak self-heating powers of massive relativistic bodies at their shortest interaction proximities, when confined emission-absorption of paired gravitational waves and emission of outward EM waves are maximized. Thermo-gravimechanical energy conservation of free accelerated bodies (with chiral masses) means that their paired gravitons (with chiral symmetry) are converted into heat locally without outward gravitational waves for distant interferometers.

Local confinement of paired GR waves can be traced quantitatively for the binary pulsar PSR B1913+16, because this object was well studied for more than   30 years \cite {Hul}. Poynting wave flows from oscillating electric charges $q$ are known from Maxwell's electrodynamics. The chiral amplitude of two unfolded gravitational waves is proportional to $2{\cal E}$. Therefore, the chiral intensity (of locally confined  vector waves with zero-balanced Poynting flows) is proportional to $4{\cal E}^2$. 
   One can replace $q^2$ in EM radiation flows with $G(2{\cal E}/ c^2)^2$ for exact computation of the radiation self-heating by paired vector gravitons.  The angular intensity of chiral gravitational waves  and the integral heat balance for equal and constant energy-charges ${\cal E}$ on the opposite circular orbit of radius `a' can be computed from the readily available electrodynamic solutions \cite {Bat},
   \begin {equation}
   \cases {
   {dI_{{_G}{_R}}}/{d\Omega} = {G (2{\cal E} /c^2)^2}  { 2 a^4\omega^6 sin^2\theta(1+cos^2\theta)} / {\pi c^5}\cr\cr
   128 G{\cal E}^2 a^4 \omega^6 /5c^9 = 
     d<N>/dt = - d <{ K} + { U_o}>/dt.\cr
      }
   \end {equation}
   
     The non-relativistic motion of such a symmetrical two-body gyroscope with ${\cal E} \approx mc^2$ and     $\omega = (Gm   /4a^3)^{1/2} = 2\pi/P$ complies with  the virial theorem, which directly relates the system kinetic energy increase to the internal energy increase,   $ mdv^2/ dt = d N/dt $. 
     This energy balance for mechanical and thermodynamical degrees of freedom explains decays of pulsar periods P under the `dissipation' decrease of the mutual proximity $`2a'$ of rotating stars. Thermo-gravitational self-organization of two radial bodies with constant energy-charges, ${\cal E}_1 = const$ and ${\cal E}_2 = const$, into one radial body (with coalesced energies ${\cal E}_1 +{\cal E}_2$ around one center of radial symmetry when $`a'\rightarrow 0$) is not accompanied by outward gravitational waves, but only by electromagnetic signals due to continuous gamma decay of GR charges with internal energy (and temperature for the black-body radiation).
         
        Slow velocities $v = 2\pi a/P < c$ result from (2) in the well known period decay
        $ dP/dt = -(48 \pi/5 c^5) (4\pi Gm/P)^{5/3}$ $ = - 3.4 \times 10^{-12} (m \cdot 1 h / M_{Sun} P)^{5/3}$  for symmetrical pulsars. This `electrodynamic' result for confined vector gravitons matches the Hulse-Taylor pulsar data, $dP/dt = (-2.422 \pm 0.006)\times 10^{-12}$, $P = 7.75 h$, after mass asymmetry corrections for  
        $m \approx 1.4 M_{Sun} $.  Paired (chiral) gravitons carry a zero Pointing vector and, therefore, the two-body gravitational interaction cannot be intercepted or screened by third bodies. However, such chiral energy-information exchanges instantly increase heat contents of constant interacting charges of the two-body system that can be predicted and tested in practice. 
        
            The strong field energy $U_o = - G{\cal E} {\cal E}_{_M}/c^4 r = - {\cal E}r_o/r $  in (1) for cold neutron stars (without self-heating) next to a giant massive companion, ${\cal E } / {\cal E}_{_M} \equiv G {\cal E}/ r_o c^4  = const << 1 $ would correspond to the classical atomic orbit $ 1/r = A + B cos (k \phi) $, with $ A = {\cal E}^2 r_o/c^2 k^2 L^2$,   $B = [ {\cal E}^2c^{-2}k^{-4}L^{-2} - m^2c^2k^{-2}L^{-2}]^{1/2}$, $k = (1 - {\cal E}^2 r^2_o c^{-2}L^{-2})^{1/2}$, and $L^2 = ({\bf r} \times {\bf P})^2 = conts$. However, real gravitational orbits are not steady due to the continuous heat accumulation $I_{GR} = d<N>/dt > 0 $ from chiral gravitational waves of the accelerated constant charge ${\cal E}$,
                               \begin {eqnarray}
           \frac {8G {\cal E}^2 <{\dot {\bf v}}^2> } {3 c^7}  \approx \frac {d<N>}{dt}  = - \frac  {d<K+U_o>}{dt}    \approx - \frac {d <mc^2 {\sqrt {1-v^2c^{-2}}}>}{dt},  
                      \end {eqnarray}
                      even in the absence of net EM radiation losses.  
   Here we approximate self-heating of distant neutron stars by the dipole radiation, $I_{GR} \approx 8G({\ddot {\bf p}})^2 /3c^3$ and  ${{\bf p}} (t') \equiv  \int {\bf r'} \rho ({\bf r}', t') dV' \approx {\bf r}' {\cal E}/ c^2$, of paired GR charges through the Larmor radiation limit for non-relativistic electric charges.   The gravitational wave damping for equal binaries is less efficient than  for the circular motion of one partner around a static supermassive center. For example,  non-relativistic circular orbits in (3) yield a quite rapid period decay
            \begin {equation}      
             \frac {dP} {dt} = -\frac { 32\pi^2 Gm} {c^3 P} = - 4.3\times 10^{-7} (m\cdot 1h / M_{Sun} P).
             \end {equation}
              Circular orbits correspond to a steady heat release and cannot provide periodic gamma outbursts (for distant observations). Neutron stars on prolonged elliptic orbits around suppermassive companions can generate heat pulses with X-rays periodically.   Such relativistic objects with predictable period decays can test (4) for paired (confined) gravitational waves.     
        Analysis of dynamical Wein's displacements in outburst spectra could also verify their non-stationary thermal  origin from oscillating accelerations ${\dot {\bf v}}^2$ in the chiral  heat intensity (3). This may shed some light on the vanishing Poynting vector for chiral interactions (of inert and live matter) in non-empty energy space. 
        
        The self-generated heat from local `absorption' of `just emitted' anisotropic gravitational waves  warms free falling matter and modifies a Keplerian motion of stars toward their spatial coalesce under the strict energy conservation. 
        A radial discrepancy between a galaxy luminosity and a gravitational energy-matter density is not a violation of the energy-to-energy gravitational law, but a confirmation of the universal energy balance (1).  The gravitational self-heating and net  gamma outcome/income  of thermobalanced energy-charges in the energy-to-energy gravitation should be counted for exact geodesic curves of free astronomical bodies and their satellites.
        
{}

\end {document}